# Real Time Data Warehouse

## Syed Ijaz Ahmad Bukhari

Real Time Data Warehouse (RTDW) is a simulation of working of human brain. Every human brain consists of approximately one billion neurons which pass data in the shape of signals to each other via synaptic connections (about thousand trillion). The brain continuously receives sensory information; refresh its data repository and responds differently in different situations. The memory capacity of a brain is around million gigabytes which gives human wonderful power to think, speak, dream, analyze, and execute plans. RTDW also aims answering queries, analyzing present trends and forecasting future outcomes in real time.

Key developments in data warehousing started in 1960s when General Mills and Dartmouth College used terms dimensions and facts in a joint research project. In 1970s ACNielsen and IRI used the term data marts for retail sales [1]. Operational systems and analytical processes were first time differentiated by MIT researchers. Bill Inmon also defined and discussed the term Data Warehouse (DW). In 1975, Sperry Univac Introduced database management and reporting system: MAPPER (MAintain, Prepare, and Produce Executive Reports) that is reported as the first platform specifically designed for building Information Centers (a forerunner of contemporary Enterprise Data Warehousing platforms). Development of the "Business Data Warehouse" by IBM researchers Barry Devlin and Paul Murphy in 1980 brought a fundamental change in architectural design of database systems [2]. In 1983, Teradata introduced DBMS specifically designed for decision support systems. In 1984, Metaphor Computer Systems released a package comprising hardware/software and GUI named Data Interpretation System (DIS) for business users.

The term Business Data Warehouse was first time published in IBM Systems Journal by Barry Devlin and Paul Murphy in their article titled "An Architecture for a Business and Information System". Ralph Kimball, introduced the DBMS, "Red Brick Warehouse" specifically designed for data warehousing in 1990. In 1991, Bill Inmon's firm Prism Solutions introduced "Prism Warehouse Manager" software for developing a data warehouse. He also published the book titled *"Building the Data Warehouse"* [3]. Ralph Kimball published the book "*The Data Warehouse Toolkit*" in 1996 [4]. In 2000, Daniel Linstedt released the "Data Vault" enabling Real Time Data Warehousing.

Bill Inmon has described Data Warehousing as a subject-oriented, integrated, time-variant, and non-volatile collection of data in support of management's decision-making process [5]. DW is primarily designed to answer queries, analysis and forecasting. It receives data from various sources namely operational data through transactions from hierarchical, distributed, and network databases, departmental file systems, data stored on PCs and servers, internets, etc. This data is historic and non-volatile in nature and is refreshed.

Data from various sources is integrated, cleaned, transformed, and cataloged before uploading to the DW. DW is subdivided into Data Marts (DM) for faster access and ease of use. The main aim of DW is to facilitate the decision making process.

DW and Online Transaction Process (OLTP) systems are different as DWs are not in Third Normal Form which is a general requirement for OLTP systems. DW systems are designed for ad hoc query processing whereas OLTP systems provide fast transaction processing. Other differences are as following:

| Operations | DW | OLTP |
|---|---|---|
| Workload | Ad hoc queries | Predefined Operations |
| Data Modifications | Regularly Updated by Extract Transform and Load Operations | Always Updated |
| Schema Design | De-normalized or partially De-normalized schemas | Normalized Schemas |
| Typical Operations | Scans thousands or millions of records for Query Processing | Accesses small number of records |
| Historical Data | Stores data of many months or years | Stores current and operational data |
| Support | Historical Analysis | Current Transaction |

Ralph Kimball has described Nine Step Methodology to design DW which includes Choosing the Process, Choosing the Grain, Identifying and Conforming the Dimensions, Choosing the Facts, Storing Pre-calculations in the Fact Table, Rounding out the Dimensions, Choosing the Duration of the Database, Tracking Slowly Changing Dimensions, and Deciding the Query Priorities and Query Modes [5]. A typical structure of a DW is as following:

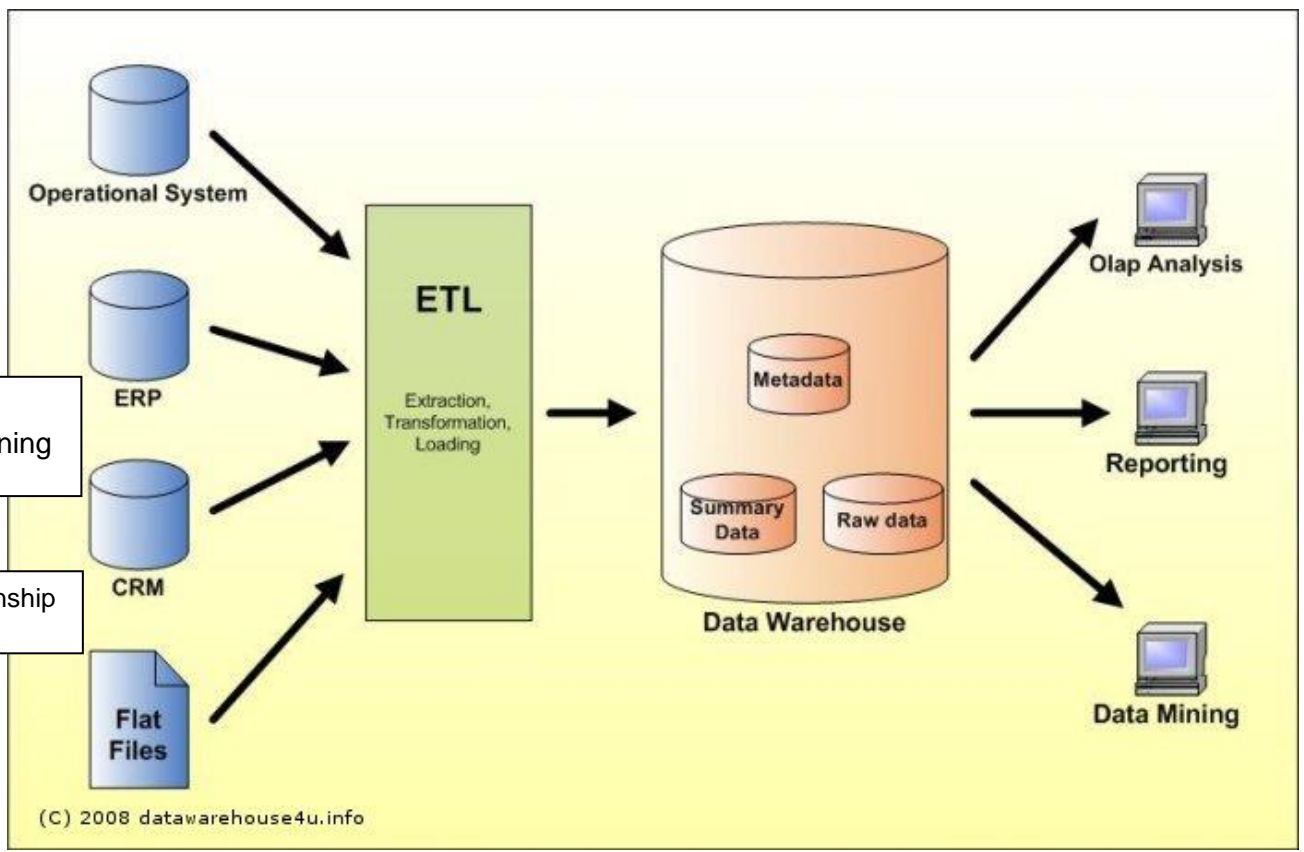

DW techniques have evolved with innovations in technology. In Offline Operational DW the data is added to DW repository on regular basis from operational data bases. However, this data is integrated, cleaned and transformed before entering into DW. In Offline DW, the data is stored in a specific data structure to facilitate reporting and answering queries. In On Time DW the data is updated on performance of each transaction. In Online Integrated DW data is received from many sources which provide facility of moving across different systems for reporting and answering queries. In RTDW, DW is updated with influx of new data. For example, in an Air Ticketing System, the data is updated as a ticket is sold.

DWs contain historic and static data. It is very important to note that how data is processed at Staging Area. Data flows from Operational Data Stores as well as other sources to the Staging Area, where it is cleaned. It is like the Oil Refinery which gathers crude oil from different sources for refinement and distributes it to various sale points for selling to the consumers. The quality of this data could also be measured by metric like Signal to Noise Ratio. Optimization methods could improve the quality of data and subsequently the queries, but it is not the only option. Thus, flow of data is an important aspect like flow of water in waterfalls. Real time Data Flow will, therefore transform a DW to RTDW.

DWs are working well to answer ad hoc queries. But, the question arises how effective are these systems when dealing with financial markets, banking transactions, reservation of seats, etc. Today's, business environment makes static DWs less valuable. The DW analysts and designers should decide how quickly and frequently they can provide more current information. In some cases, security and accuracy of this data will also come in play, particularly where information of national interest are involved.

RTDW implementation is an answer to above challenges that will help the user, employees and management to access the most current data for preparing reports and summaries, analysis for decision making, and forecasting future trends. DWs provide consistent and relevant data from sources as data is integrated, cleaned and transformed before entering into DW. However, DW takes more time in creation and operation. Continuous up gradation of technology and software in line with future data requirements are also required. Security of the data is another consideration, as data is received from many sources including internets.

Justin Langseth has discussed the challenges in developing RTDW that include Enabling Real-time ETL, Modeling Real-time Fact Tables, and OLAP Queries vs. Changing Data, Scalability & Query Contention, and Real-time Alerting [6]. He has also suggested some solutions to these also which are being discussed in preceding paras.

Extraction, Transformation, and Loading (ETL) of cleaned data is the most difficult step in RTDW. ETL is generally a batch process and users are not able to access the data during loading process. Therefore, the DWs are refreshed in night hours or when less number of transactions is expected. In RTDW, this luxury of downtime is not available. However, specialized ETL tools are available in the market that provide real time or near real time data loading. One of the approach may be that frequency of the data loading may be increased, for example from week to daily and daily to hourly and hourly to half hourly, etc. This approach does not involve many modifications or heavy costs, but user will be getting more fresh data. Second approach could be inserting/ updating data continuously in DW Fact tables or by inserting data into separate fact tables in a real-time partition. DataMirror and MetaMatrix real time data loading packages have been designed for this specific purpose [6]. However, these approaches may have scalability problem as the DW have huge data.

The scalability problem could be overcome with Trickle and Flip approach. The data could be fed into Staging Tables instead of actual DW Fact tables. These Staging Tables should be designed exactly same as DW Fact tables and should contain a copy of the data for the current duration. The data from Staging Tables to DW Fact tables be duplicated periodically. But, this allow need temporarily pause of the OLAP server during flip period. The approaches discussed so far need taking extra load to underlying databases. Another approach, External Real-time Data Cache (RTDC) stores real time data outside the DW not affecting the resources of traditional DW. RTDC is a dedicated Database server performing ETL on real time data. Angara, Cacheflow, Kx, TimesTen, and InfoCruiser provide In-memory Database (IMDB) for real time data cache. In this approach, all real time data is added to the cache regardless of the type of database being used.

Modeling Real Time Fact Tables is second challenge, because the time dimensions may behave strangely with continuous changes in time. The real challenge is that how data is stored and what are its links with rest of the data modeling. One approach is to store the real time data in separate DW Fact tables and query tools can be set up to move across to get historical as well as real time data. But, this is a complex and very difficult option to engineer from design and administrative perspectives [1]. Alternatively, real time data can be stored in different tables from historical data, but in same table structure. These should look like one logical table and will solve many problems associated with separate storage option. However, one should be careful that query tools do not return old cache results. To overcome this problem, external data cache can be used and some additional tables such as look up tables can be integrated with historical data for querying and analysis.

Traditional DWs contain historic static and unchanging data; therefore these produce correct reports, summaries and analysis. Duration of query processing ranges from fraction of seconds to many hours. Now, the problem with RTDWs is that the data may change during the course of OLAP and query processing, thus producing inaccurate reports and analysis. To meet this challenge, Near-real-time ETL approach or the Trickle & Flip approach with a relatively long cycle time be used. However, OLAP server be paused during load or flip process so that user can get near real time reports and queries. In order to get real time analysis and reports, real time data can be stored separately and can be merged using Just in Time Information Merging (JIM) approach.

The issue of scalability and query contention is more complex for RTDW. This issue may be resolved by adding more memory and faster processors. But, this is a short term options, because it is difficult to say how much memory and how fast the processor and will these be effective for most of the time. External Real Time Data Cache may resolve this issue by routing all real time data to a separate cache database and join real time and historic information to answer queries and analysis. This approach will also not produce satisfactory results for Complex Analytic reports. Main problem with RTDW is to provide accurate and complex reports and OLAP queries which involve millions of transaction per second and being accessed by thousands of users concurrently. It can be achieved by placing historic data in DW and real time information in an external cache. JIM Request Analyzer will pre-process the real time data to determine what components are really required for query processing. JIM Data Imager will take a snapshot of the required real time data and will load this snapshot into temporary tables for processing the query along with historic data of DW [6]. A variant of this approach is Reverse JIM. In this approach needed historic information is loaded temporary into data cache and query is run in the data cache. The problem with this approach is that data cache may overflow. Thus, the best option is to use both options and query should decide to select which option basing on the best direction of the data flow.

Real Time Alerting operates on occurrence of an event or on completion of scheduled task. Alerts are triggered to the users after few minutes or hours. Presently, DW alerting packages trigger alerts to every 1, 5, and 15 or 30 minutes which give a near real time solution [6]. However, it should be ensured that same alert is not sent continuously again and again in same alerting cycle, so that a threshold problem does not occur. Therefore, in real time alerting, the triggering schedule needs to be minimized. But, it may cause errors or missing alerts as some processes may start before the previous has not ended. Moreover, for a real time system like Stock Market having thousands of users, millions of transactions and alerts coming with every tick of second will face great difficulties. However, it is important that in any RTDW, alert triggering is managed in a manner that is acceptable to the user.

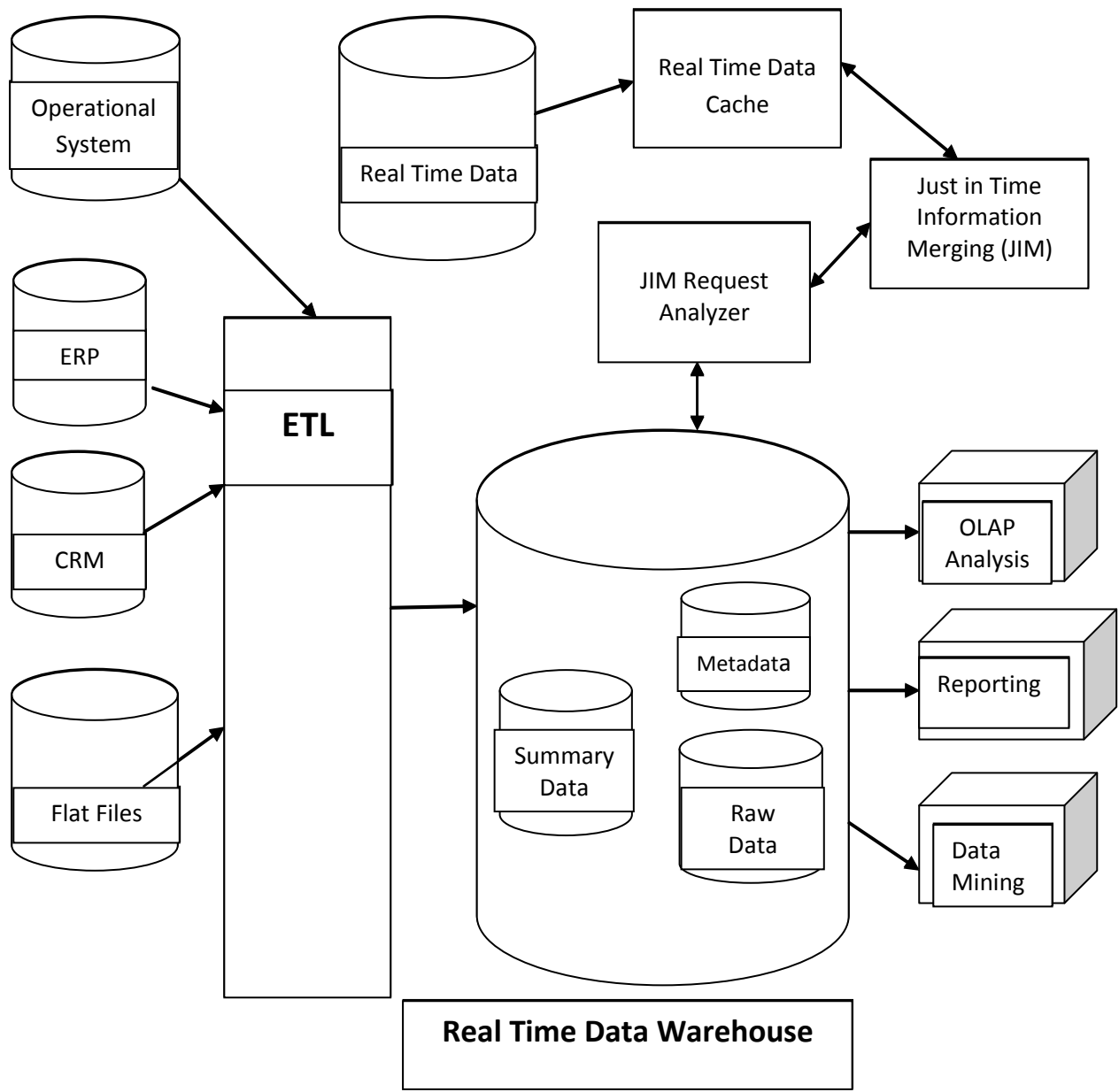

**Real Time Data Warehouse**

## Conclusion

RTDW is a reality and present technologies are supporting it with some problems. Real time or near real time reporting, querying, analysis and alerting is a tradeoff between cost and user requirements. The benefits of RTDW are becoming increasingly clear to the users and it is expected that problems discussed in this paper will be eradicated with advancement in technology.